\documentclass[11pt]{article}
\usepackage{amsmath,amssymb}
\usepackage[dvips]{graphicx}
\usepackage{txfonts}

\newtheorem{theorem}{Theorem}[section]
\newtheorem{definition}{Definition}[section]
\newtheorem{lemma}{Lemma}[section]

\newtheorem{claim}{Claim}[section]

\newenvironment{proof}{\par \noindent
            {\bf Proof. \hspace{2mm}}}{\hfill$\Box$ \vspace*{3mm}}
\newenvironment{proofof}[1]{\vspace*{5mm} \par \noindent
         {\bf Proof of #1.\hspace{2mm}}}{\hfill$\Box$ \vspace*{3mm}}

\def\bra#1{\langle #1 |}
\def\ket#1{| #1 \rangle}
\def\kb#1#2{\ket{#1}\bra{#2}}
\def\fixbitE#1#2#3#4{{#1}_{(#3,#4)}={#2}_{(#3,#4)}}
\def\fixbitN#1#2#3#4{{#1}_{(#3,#4)}\ne {#2}_{(#3,#4)}}
\def\udn#1{#1\in\{0,1\}^n}

\begin{document}
\begin{center}
{\Large\bf
Universal Test for Quantum One-Way Permutations}\\[7mm]
\large
Akinori Kawachi$^{\dag,\S}$ \ \ 
Hirotada Kobayashi$^\ddag$ \ \ 
Takeshi Koshiba$^{\dag,\P}$ \ \ 
Raymond H. Putra$^{\dag,\S}$\\[7mm]
\normalsize
\begin{tabular}{l@{\hskip 1mm}l}
$^\dag$ & \sl Quantum Computation and Information Project,\\
& \sl ERATO, Japan Science and Technology Agency\\
& 406 Iseya-cho, Kawaramachi-Marutamachi, Kamigyo-ku, Kyoto 602-0873, Japan.\\
& \tt \verb+{kawachi,koshiba,raymond}@qci.jst.go.jp+\\[2mm]
$^\ddag$ & \sl Quantum Computation and Information Project,\\
& \sl ERATO, Japan Science and Technology Agency\\
& 5-28-3 Hongo, Bunkyo-ku, Tokyo 113-0033, Japan.\\
& \tt\verb+hirotada@qci.jst.go.jp+\\[2mm]
$^\S$ & \sl Graduate School of Informatics, Kyoto University\\
& Yoshida-Honmachi, Sakyo-ku, Kyoto 606-8501, Japan.\\[2mm]
$^\P$ & \sl Secure Computing Laboratory, Fujitsu Laboratories Ltd.\\
& 4-1-1 Kamikodanaka, Nakahara-ku, Kawasaki 211-8588, Japan.\\
\end{tabular}
\end{center}

\vspace*{1mm}

\begin{abstract}
The next bit test was introduced by Blum and Micali 
and proved by Yao to be a universal test 
for cryptographic pseudorandom generators. On the other hand,
no universal test for the cryptographic one-wayness of
functions (or permutations) is
known, though the existence of cryptographic pseudorandom generators
is equivalent to that of cryptographic one-way functions.
In the quantum computation model, Kashefi, Nishimura and Vedral
gave a sufficient condition of (cryptographic)
quantum one-way permutations and conjectured that the condition
would be necessary. In this paper, we affirmatively settle
their conjecture and complete 
a necessary and sufficient for quantum one-way permutations. 
The necessary and sufficient condition can be regarded
as a universal test for quantum one-way permutations, since the
condition is described as a collection of stepwise tests similar to
the next bit test for pseudorandom generators.
\end{abstract}

\section{Introduction}
One-way functions are functions $f$ such that,
for each $x$,  $f(x)$ is efficiently computable
but, only for a negligible fraction of $y$, $f^{-1}(y)$ is
computationally tractable. While the modern cryptography
depends heavily on one-way functions,
the existence of one-way functions is one of the most important open
problems in theoretical computer science.
On the other hand, Shor \cite{shor97} showed that famous candidates
of one-way functions such as the RSA function or the discrete logarithm function
are no longer one-way in the quantum computation model.
Nonetheless, some cryptographic applications based on quantum one-way functions
have been considered (see, e.g., \cite{ac02,dms00}).

As a cryptographic primitive other than one-way functions,
pseudorandom generators have been studied well. 
Blum and Micali \cite{bm84} proposed how to construct
pseudorandom generators from one-way permutations
and introduced the next bit test for pseudorandom generators.
(They actually constructed a pseudorandom generator
assuming the hardness of the discrete logarithm problem.)
Since Yao \cite{yao82} proved that the next bit test is a universal test
for pseudorandom generators, the Blum--Micali's construction paradigm
of pseudorandom generators from one-way permutations was accomplished.
In the case of pseudorandom generators based on one-way permutations,
the next bit unpredictability can be proved by using the hard-core
predicates for one-way permutations. After that, Goldreich and Levin \cite{gl89}
showed that there exists a hard-core predicate for any one-way function
(and also permutation) and H{\aa}stad {\em et al.} \cite{hill99}
showed that the existence of pseudorandom generators is equivalent 
to that of one-way functions.

Yao's result on the universality of the next bit test assumes that any
bits appeared in pseudorandom bits are computationally unbiased.
Schrift and Shamir \cite{ss93} extended Yao's result to the biased case
and proposed universal tests for nonuniform distributions.
On the other hand, 
no universal test for the one-wayness of a function
(or a permutation) is known, although pseudorandom generators and
one-way functions (or permutations) are closely related.

In the quantum computation model,
Kashefi, Nishimura and Vedral \cite{knv02} gave a 
necessary and sufficient condition for the existence of {\em worst-case\/}
quantum one-way permutations. They also considered the {\em cryptographic
(i.e., average-case)\/}
quantum one-way permutations and gave a sufficient condition
of (cryptographic) quantum one-way permutations. They also conjectured
that the condition would be necessary.
Their conditions are based on the efficient implementability of 
reflection operators about some class of quantum states. 
Note that the reflection operators
are successfully used in the Grover's algorithm \cite{gro96}
and the quantum amplitude amplification technique \cite{bhmt00}.
To obtain a sufficient condition of cryptographic quantum one-way
permutations, a notion of ``pseudo identity'' operators was
introduced \cite{knv02}.
Since the worst-case hardness of
reflection operators is concerned with the worst-case hardness of the
inversion of the permutation $f$, we need some technical tool
with which the inversion process of $f$ becomes tolerant of
some computational errors in order to obtain a sufficient
condition of cryptographic quantum one-way permutations. 
Actually, pseudo identity operators
permit of {\em exponentially\/} small errors during
the inversion process \cite{knv02}.

In this paper, we complete a necessary and sufficient condition
of cryptographic quantum one-way permutations 
conjectured in \cite{knv02}.
We incorporate their basic ideas with a probabilistic argument
in order to obtain a technical tool to permit of 
{\em polynomially\/} small errors during the inversion process. 
Roughly saying, pseudo identity operators are close to
the identity operator in a sense. The similarity is defined
by an intermediate notion between the statistical distance and
the computational distance. 
In \cite{knv02}, 
it is ``by upper-bounding the similarity'' that
the sufficient condition of cryptographic quantum one-way permutations 
was obtained.
By using a probabilistic argument,
we can estimate the expectation of the similarity and then handle
polynomially small errors during the inversion of the permutation $f$.

Moreover, the necessary and sufficient condition of
quantum one-way permutations can be regard as
a universal test for the quantum one-wayness of permutations.
To discuss universal tests for the one-wayness of permutations, we briefly
review the universality of the next bit test for pseudorandom generators.
Let $g(x)$ be a length-regular deterministic function such that
$g(x)$ is of length $\ell(n)$ for any $x$ of length $n$.
The universality of the next bit test says that we have only to check 
a collection of stepwise polynomial-time tests $T_1,...,T_{\ell(n)}$ instead
of considering all the polynomial-time tests that try to
distinguish the truly random bits from output bits from $g$,
where each $T_i$ is the test whether, given the $(i-1)$-bits 
prefix of $g(x)$ (and the value of $\ell(|x|)$),
the $i$-th bit of $g(x)$ is 
predictable or not  with probability non-negligibly higher than 1/2.
Our necessary and sufficient condition of
quantum one-way permutations says that the quantum one-wayness of a given 
permutation $f$ can be checked by a collection of
stepwise tests $T_1',...,T_n'$ instead of considering all the tests
of polynomial-size quantum circuit, where each $T_i'$ is the test whether,
given some quantum state $q_{i-1}$ that can be defined
by using the $(i-1)$-bits prefix of $f(x)$, 
some other quantity $t_i$ is computable with polynomial-size quantum circuit
or not and the next state $q_i$ can be determined from $q_{i-1}$ and $t_i$.
In this sense, our universal test for quantum one-way permutations
is analogous to the universal test (i.e., the next bit test)
for pseudorandom generators.

\section{Preliminaries}
We say that a unitary operator (on $n$ qubits) is {\em easy\/} if
there exists a quantum circuit implementing $U$
with polynomial size in $n$ and
a set $\cal F$ of unitary operators is {\em easy\/} if 
every $U\in\cal F$ is easy.
Throughout this paper, we assume that $f:\{0,1\}^\ast \rightarrow \{0,1\}^\ast$
is a length-preserving permutation unless otherwise stated. 
Namely, for any $x\in\{0,1\}^n$, $f(x)$ is an $n$-bits string and the set
$\{f(x): x\in\{0,1\}^n\}$ is of cardinality $2^n$ for every $n$.
First, we mention some useful operators in describing the previous and our
results. The tagging operators $O_j$ are defined as follows:
\[ O_j\ket{x}\ket{y} = 
\begin{cases}
- \ket{x}\ket{y} & {\rm if~} \fixbitE{f(y)}{x}{2j}{2j+1}\\
\ket{x}\ket{y} & {\rm if~} \fixbitN{f(y)}{x}{2j}{2j+1}
\end{cases}
\]
where $y_{(i,j)}$ denotes the substring from the $i$-th bit
to the $j$-th bit of the bit string $y$. 
Note that these unitary operators $O_j$ are easy.
Next, we consider the 
reflection operators $Q_j(f)$ as follows:
\[
Q_j(f) = \sum_{\udn{x}}\kb{x}{x}\otimes (2\kb{\psi_{j,x}}{\psi_{j,x}}-I)
\]
where
\[
\ket{\psi_{j,x}} = \frac{1}{\sqrt{2^{n-2j}}}
	\sum_{y:\fixbitE{f(y)}{x}{1}{2j}} \ket{y}.
\]
(See Fig.\ 1 for the reflection operator.)
We sometimes use the notation $Q_j$ instead of $Q_j(f)$.
\begin{center}
\mbox{}\\
\scalebox{0.4}{\includegraphics{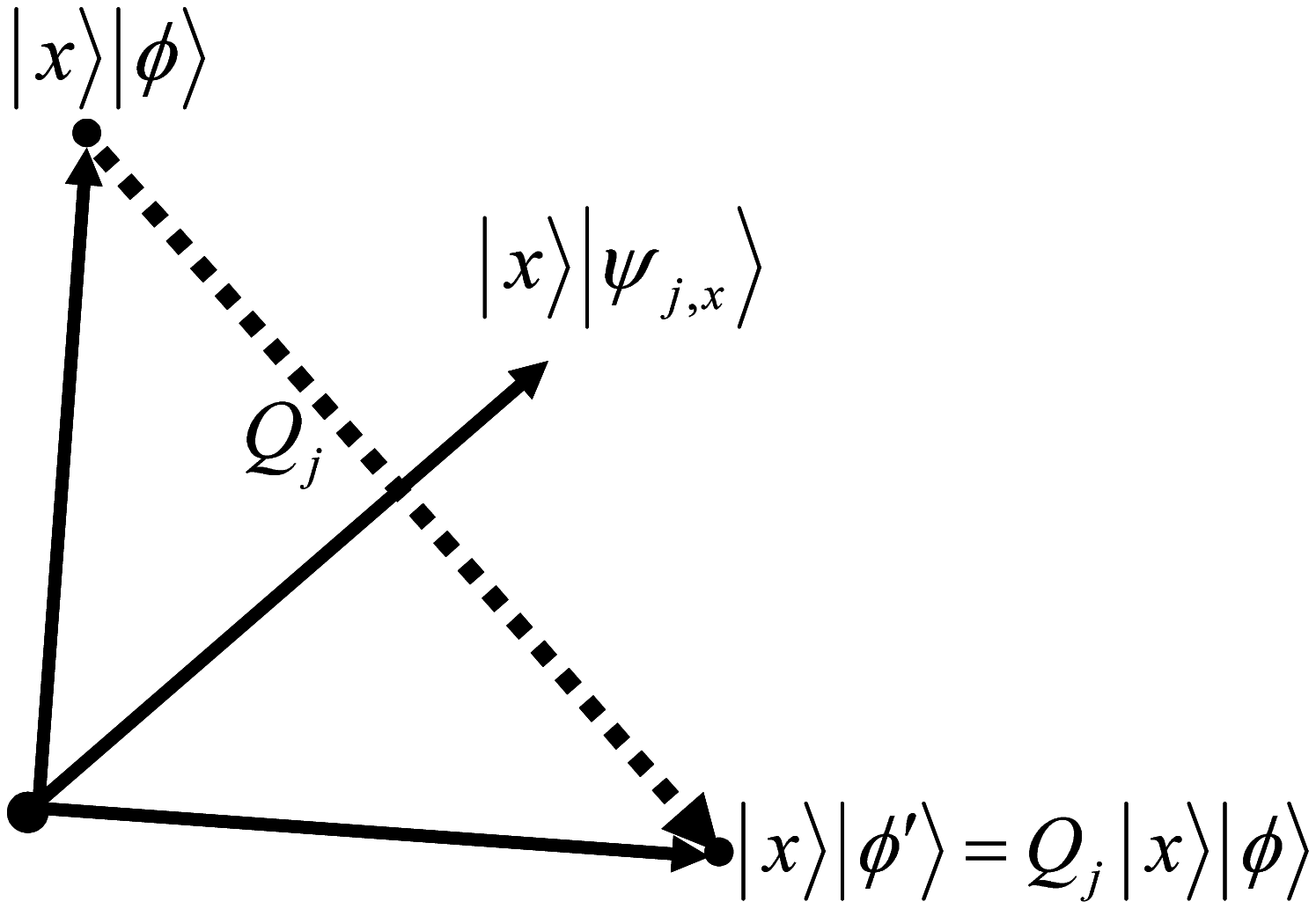}}\\
Fig.~1: Reflection operator\\
\mbox{}
\end{center}
Actually, these reflection operators are somewhat special for our purpose.
In general, reflection operators are commonly and
successfully used in the Grover's algorithm \cite{gro96} and 
the quantum amplitude amplification technique \cite{bhmt00}.

\begin{theorem}\label{thm:wc}{\rm
(Kashefi, Nishimura and Vedral \cite{knv02})}
Let $f:\{0,1\}^n\rightarrow\{0,1\}^n$ be a permutation. Then
$f$ is worst-case quantum one-way if and only if the set
${\cal F}_n = \{Q_j(f)\}_{j=0,1,...,\frac{n}{2}-1}$ of unitary operators
is not easy.
\end{theorem}

As a part of the proof of Theorem \ref{thm:wc}, Kashefi, Nishimura and
Vedral \cite{knv02} give a quantum algorithm 
(we call Algorithm {\sf INV} in what follows)
computing $f^{-1}$ by using unitary operators $O_j$ and $Q_j$.
The initial input state to {\sf INV} is assumed to be
\[ \frac{1}{\sqrt{2^n}}\ket{x}\sum_{\udn{y}}\ket{y}, \]
where {\sf INV} trys to compute $f^{-1}(x)$. Then
{\sf INV} performs the following steps:
\begin{quote}
{\bf foreach} $j=0$ to $\frac{n}{2}-1$\\
\hspace*{7mm} ({\sf step W.j.1}) Apply $O_j$ to the first and the second registers;\\
\hspace*{7mm} ({\sf step W.j.2}) Apply $Q_j$ to the first and the second registers.
\end{quote}
After each step, we have the following:
\begin{eqnarray*}
\mbox{(the state after {\sf step W.j.1})} & = &
\frac{2^j}{\sqrt{2^n}}\ket{x}\left( \sqrt{2^{n-2j}}\ket{\psi_{j,x}} -
2\sum_{y:\fixbitE{f(y)}{x}{1}{2j+2}}\ket{y}\right).\\
\mbox{(the state after {\sf step W.j.2})} & = &
\frac{2^{j+1}}{\sqrt{2^n}}\ket{x}\sum_{y:\fixbitE{f(y)}{x}{1}{2j+2}}\ket{y}.
\end{eqnarray*}

Before reviewing a known sufficient condition of cryptographic quantum one-way permutations,
we define two types of cryptographic ``one-wayness'' in the
quantum computational setting.

\begin{definition}\rm
A permutation $f$ is {\em weakly quantum one-way\/} if the following conditions
are satisfied: 
\begin{enumerate}
\item $f$ can be computed by a polynomial size quantum circuit (and 
whenever inputs are classical the corresponding outputs must be classical).
\item There exists a polynomial $p(\cdot)$ such that for every polynomial size
quantum circuit $A$ and all sufficiently large $n$'s,
\[ \Pr[A(f(U_n))\ne U_n] > \frac{1}{p(n)}, \]
where $U_n$ is the uniform distribution over $\{0,1\}^n$.
\end{enumerate}
\end{definition}

\begin{definition}\rm
A permutation $f$ is {\em strongly quantum one-way\/} if the following conditions
are satisfied: 
\begin{enumerate}
\item $f$ can be computed by a polynomial size quantum circuit (and 
whenever inputs are classical the corresponding outputs must be classical).
\item For every polynomial size quantum circuit $A$ and every
polynomial $p(\cdot)$ and all sufficiently large $n$'s,
\[ \Pr[A(f(U_n))= U_n] < \frac{1}{p(n)}. \]
\end{enumerate}
\end{definition}

As in the classical one-way permutations, we can show that the existence
of weakly quantum one-way permutations is equivalent to that
of strongly quantum one-way permutations
(see, e.g., \cite{goldreich}).
Thus, we consider the weakly quantum
one-way permutations in this paper. 
While Theorem \ref{thm:wc} is a necessary and sufficient condition
of {\em worst-case\/} quantum one-way permutations,
Kashefi, Nishimura and Vedral \cite{knv02} also gave a sufficient
condition of {\em cryptographic\/} quantum one-way permutations by using 
the following notion.

\begin{definition}\rm
Let $d(n)\ge n$ be a polynomial in $n$ and $J_n$ be a $d(n)$-qubit
unitary operator. $J_n$ is called $(a(n),b(n))$-pseudo identity
if there exists a set $X_n\subseteq \{0,1\}^n$ such that
$|X_n|/2^n \le b(n)$ and for any $\udn{z}\setminus X_n$
\[ | 1 - (\bra{z}_1\bra{0}_2)J_n(\ket{z}_1\ket{0}_2)| \le a(n), \]
where $\ket{z}_1$ is the $n$-qubit basis state for each $z$
and $\ket{0}_2$ corresponds to the ancillae of $d(n)-n$ qubits.
\end{definition}

The closeness between a pseudo identity operator and the identity operator
is measured by a pair of parameters $a(n)$ and $b(n)$. The first
parameter $a(n)$ is a measure of a statistical property and the second
one $b(n)$ is a measure of a computational property. Note that we do not
care where each $z\in X_n$ is mapped by the pseudo identity operator $J_n$.
While we will give a necessary and sufficient condition of quantum one-way permutations
by using the notion of pseudo identity, 
we introduce a new notion, which
may be helpful to understand intuitions of our and previous conditions,
in the following.

\begin{definition}\rm
Let $d'(n)\ge n$ be a polynomial in $n$ and $P_n$ be a $d'(n)$-qubit
unitary operator. $P_n$ is called $(a(n),b(n))$-pseudo reflection
(with respect to $\ket{\psi(z)}$)
if there exists a set $X_n\subseteq \{0,1\}^n$ such that
$|X_n|/2^n \le b(n)$ and for any $\udn{z}\setminus X_n$
\[ \left| 1 - \biggl(\bra{z}_1\bra{w}_2
\Bigl(\sum_{\udn{y}}\ket{y}\bra{y}_1
\otimes
(2\ket{\psi(y)}\bra{\psi(y)}-I)_2\Bigr)
\bra{0}_3\biggr)P_n
(\ket{z}_1\ket{w}_2\ket{0}_3)\right| \le a(n). \]
\end{definition}

The above definition of pseudo reflection operators is somewhat complicated.
Since Fig.\ 2 illustrates a geometrical intuition, it may be helpful to
understand the idea of pseudo reflection operators.
Let $J_n$ be a $d(n)$-qubit $(a(n),b(n))$-pseudo identity operator.
Then $(I_n\otimes J_n)^{\dag}(Q_j\otimes I_{d(n)-n})
(I_n\otimes J_n)$ is a $(d(n)+n)$-qubit
$(a'(n),b'(n))$-pseudo reflection operator with respect to $\ket{\psi_{j,x}}$,
where $a'(n)\le 2a(n)$ and $b'(n)\le 2b(n)$.
These estimations of $a'(n)$ and $b'(n)$ are too rough to obtain a
necessary and sufficient condition. Rigorously estimating these parameters is a main technical
issue in this paper.

\begin{theorem}\label{thm:avc}
{\rm (Kashefi, Nishimura and Vedral \cite{knv02})}
Let $f$ be a permutation that can be computed by a polynomial-size 
quantum circuit. If $f$ is not (weakly) quantum one-way, then
for any polynomial $p$ and infinitely many $n$, there exist a
polynomial $r_p(n)$ and a $r_p(n)$-qubit $(1/2^{p(n)},1/p(n))$-pseudo
identity operator $J_{n}$ such that the family of pseudo reflection
operators
\[ {\cal F}_{p,n}(f) = \{ (I_n\otimes J_{n})^{\dag}
	(Q_j(f)\otimes I_{r_p(n) -n})
	(I_n\otimes J_{n})\}_{j=0,1,...,\frac{n}{2}-1} \]
is easy.
\end{theorem}

Kashefi, Nishimura and Vedral \cite{knv02} conjectured that
the converse of Theorem \ref{thm:avc} should still hold
and proved a weaker version of the converse as follows.

\begin{theorem}
{\rm (Kashefi, Nishimura and Vedral \cite{knv02})}
Let $f$ be a permutation that can be computed by a polynomial-size 
quantum circuit. If for any polynomial $p$ and infinitely many $n$
there exist a polynomial $r_p(n)$ and a $r_p(n)$-qubit
$(1/2^{p(n)},p(n)/2^n)$-pseudo identity operator $J_n$ such that
the family of pseudo reflection operators
\[ {\cal F}_{p,n}(f) = \{ (I_n\otimes J_{n})^{\dag}
	(Q_j(f)\otimes I_{r_p(n) -n})
	(I_n\otimes J_{n})\}_{j=0,1,...,\frac{n}{2}-1} \]
is easy, then $f$ is not (weakly) quantum one-way.
\end{theorem}

\begin{center}
\mbox{}\\
\scalebox{0.6}{\includegraphics{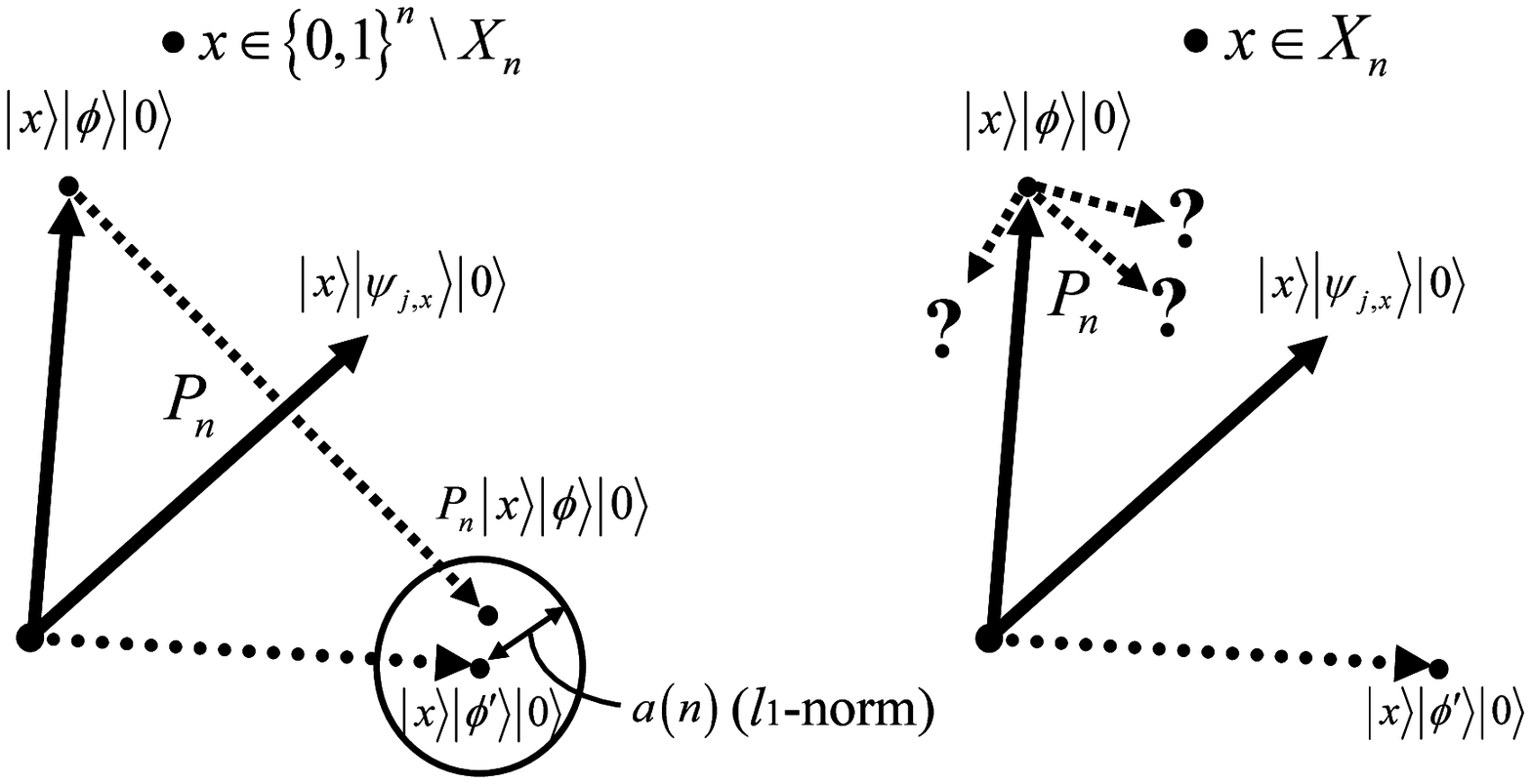}}\\
Fig.~2: Pseudo reflection operator\\
\mbox{}
\end{center}

We mention why it is difficult to 
show the converse of Theorem \ref{thm:avc}.
To prove it by contradiction,
all we can assume is the existence of a pseudo identity operator.
This means that we cannot know how the pseudo identity operator is close
to the identity operator. To overcome this difficulty, we introduce
a probabilistic technique and estimate the expected behavior
of the pseudo identity operator. 
Eventually, we give a necessary and sufficient condition of the existence
of quantum one-way permutations in terms of reflection operators.
This says that we affirmatively settle their conjecture.

\section{Necessary and Sufficient Condition of Quantum One-way Permutations}
We have a necessary and sufficient condition of cryptographic quantum one-way permutations as follows.
\begin{theorem}\label{thm:main}
The following statements are equivalent.
\begin{enumerate}
\item There exists a weakly quantum one-way permutation.
\item There exists a polynomial-time computable function $f$ satisfying that
there exists a polynomial $p$ such that for all sufficiently large $n$'s,
any polynomial $r_p(n)$ and any $r_p(n)$-qubit
$(1/2^{p(n)},1/p(n))$-pseudo identity operator $J_{n}$ such that
the family of pseudo reflection operators
\[ {\cal F}_{n,p}(f)=\{(I_n\otimes J_{n})^{\dag}
(Q_j(f)\otimes I_{r_p(n)-n})
(I_n\otimes J_{n})\}_{j=0,1,...,\frac{n}{2}-1} \]
is not easy.
\end{enumerate}
\end{theorem}

To grasp the intuition of Theorem \ref{thm:main}, Fig 3.\ may be helpful.
Theorem \ref{thm:main} can be proved as the combination of
Theorem \ref{thm:avc} and the following theorem.

\begin{theorem}\label{thm:new2}
Let $f$ be a permutation that can be computed by a polynomial-size 
quantum circuit. If for any polynomial $p$ and infinitely many $n$
there exist a polynomial $r_p(n)$ and 
a $r_p(n)$-qubit $(1/2^{p(n)},1/p(n))$-pseudo identity operator $J_{n}$ 
such that the family of pseudo reflection operators
\[ {\cal F}_{n,p}(f)=\{\tilde{Q}_j(f)\} =
\{(I_n\otimes J_{n})^{\dag}
(Q_j(f)\otimes I_{r_p(n)-n})
(I_n\otimes J_{n})\}_{j=0,1,...,\frac{n}{2}-1} \]
is easy, then $f$ is not (weakly) quantum one-way.
\end{theorem}

\begin{center}
\mbox{}\\
\scalebox{0.4}{\includegraphics{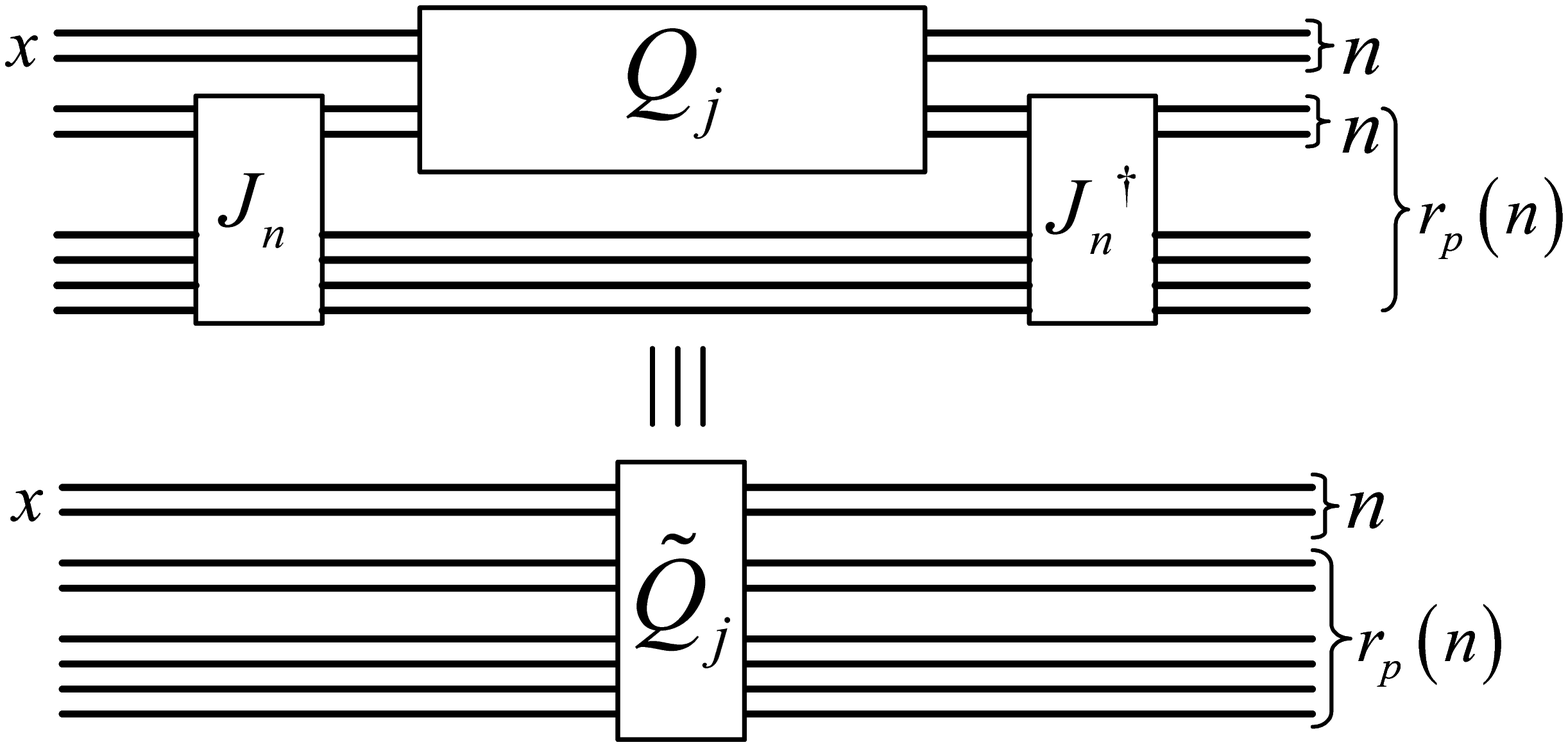}}\\
Fig.~3: Basic operations for the inversion\\
\mbox{}
\end{center}

\begin{proof}
Suppose that for any polynomial $p(n)$, 
infinitely many $n$, and some
$(1/2^{p(n)},1/p(n))$-pseudo identity operator $J_{n}$, 
the family ${\cal F}_{p,n}$ of unitary operators is easy.
Moreover, let $f$ be a weakly quantum one-way permutation. 
By a probabilistic argument, we show that a contradiction follows 
from this assumption. For more detail, we construct an efficient 
inverter for $f$ using ${\cal F}_{p,n}$ and then, if we choose 
a polynomial $p(n)$ appropriately, this efficient 
inverter can compute $x$ from $f(x)$ for a large fraction of inputs, which violates the assumption that $f$ is a weakly 
quantum one-way permutation.

We first construct a polynomial-size algorithm {\sf av-INV}
to invert $f$ by using unitary operations in ${\cal F}_{p,n}$.
Algorithm {\sf av-INV} is almost similar to Algorithm {\sf INV} except
the following change:
the operator $Q_j$ is now replaced with $\tilde{Q}_j$. 
The initial input state to {\sf av-INV} is also assumed to be
\[ \frac{1}{\sqrt{2^n}}\ket{x}_1\sum_{\udn{y}}\ket{y}_2\ket{0}_3, \]
where $\ket{z}_1$ (resp., $\ket{z}_2$ and  $\ket{z}_3$) denotes
the first $n$-qubit (resp., the second $n$-qubit and
the last $(r_p(n)-n)$-qubit) register.

Algorithm {\sf av-INV} performs the following steps:
\begin{quote}
{\bf foreach} $j=0$ to $\frac{n}{2}-1$\\
\hspace*{7mm}
({\sf step j.1}) Apply $O_j$ to the first and the second registers;\\
\hspace*{7mm} 
({\sf step j.2}) Apply $\tilde{Q}_j$ to all the registers.
\end{quote}
For analysis of Algorithm {\sf av-INV}, we use the following
functionally equivalent description. 
(Note that the following procedure may not be efficient though
the behavior is equivalent to Algorithm {\sf av-INV}.)
\begin{quote}
{\bf foreach} $j=0$ to $\frac{n}{2}-1$\\
\hspace*{7mm} 
({\sf step A.j.1}) Apply $O_j$ to the first and the second registers;\\
\hspace*{7mm} 
({\sf step A.j.2}) Apply $J_n$ to the second and third registers;\\
\hspace*{7mm} 
({\sf step A.j.3}) Apply ${Q}_j$ to the first and the second registers;\\
\hspace*{7mm}
({\sf step A.j.4}) Apply $J_n^{\dag}$ to the second and third registers.
\end{quote}

Then, we can prove the following two claims. 
\begin{claim}\label{claim:qowp}
Suppose that $f$ is a weakly quantum one-way permutation, i.e.,
 there exists a polynomial $r(n)\ge 1$ such that for every polynomial size 
quantum circuit $A$ and all sufficiently large $n$'s,
$\Pr[A(f(U_n))\ne U_n] > 1/r(n)$.
Then, there are at least $2^n(1/r(n)-1/q^2(n))/(1-1/q^2(n))$ $x$'s such that $A$ cannot compute $x$ from $f(x)$ with probability at least $1-1/q^2(n)$.
%, where $q(n) = p^{1/4}(n)/\sqrt{2n}$.
\end{claim}
\begin{claim}\label{claim:av-INV}
Let $q(n) = p^{1/4}(n)/\sqrt{2n}$.
There are at most $2^n/q(n)$ $x$'s such that Algorithm {\sf av-INV} cannot compute $x$ from $f(x)$ with probability at least $1-1/q^2(n)$.
\end{claim}

The proof of Claim \ref{claim:av-INV} is delayed and that of Claim \ref{claim:qowp} follows immediately from the definition of a weakly quantum one-way permutation
by a counting argument.

Recall that we assume that $f$ is a weakly quantum one-way permutation
at the beginning of this proof.
Now, we can set $p(n) = 4n^2(r(n)+1)^4$, that is, $q(n)=r(n)+1\ge 2$.
It follows that $(1/r(n)-1/q^2(n))/(1-1/q^2(n)) > 1/q(n)$,
which is a contradiction since {\sf av-INV} is an inverter violating 
the assumption of a weakly quantum one-way permutation $f$.
This implies that $f$ is not weakly quantum one-way.

In what follows, we present a proof of Claim \ref{claim:av-INV}
to complete the proof of this theorem.
\begin{proofof}{Claim \ref{claim:av-INV}}
From the definition of pseudo identity operators, there exists a
set $X_n\subseteq\{0,1\}^n$ with $|X_n| \le 2^n/p(n)$ such that 
for any $y\in Y_n = \{0,1\}^n\setminus X_n$,
\[ J_{n}\ket{y}_2\ket{0}_3 = \alpha_y\ket{y}_2\ket{0}_3 + \ket{\psi_y}_{23}, \]
where $\ket{\psi_y}_{23} \bot \ket{y}_2\ket{0}_3$ and
$|1-\alpha_y|\le \frac{1}{2^{p(n)}}$.

In Algorithm {\sf av-INV}, we apply $J_{n}$
before and after {\sf step A.j.3} for each $j$. 
The application of $J_{n}$ makes an error in computation of $f^{-1}$.
We call the vector $J_{n}\ket{\psi}-\ket{\psi}$ the {\em error\/}
associated to $\ket{\psi}$. To measure the effect of this error, 
we use the following lemmas. 
(Lemma \ref{lem:keep} itself was stated in \cite{knv02}.)
We note, in the sequel, the norm over vectors is Euclidean.

\begin{lemma}\label{lem:bound}
Assume that $T\subseteq S\subseteq \{0,1\}^n$. Then length $l(S,T)$ of
the error associated to the state
\[
\ket{\psi(S,T)} = \frac{1}{\sqrt{|S|}}\left(
\sum_{y\in S\setminus T}\ket{y}\ket{0} - \sum_{y\in T}\ket{y}\ket{0}\right)
\]
satisfies that
\[ l(S,T) \le 2\sqrt{\frac{|S\cap X_n|}{|S|}}+\gamma(n), \]
where $\gamma(n)$ is a negligible function in $n$.
\end{lemma}
\begin{proof}
First, we restate the property of the length of the error associated
to the state $\ket{y}\ket{0}$ which was shown in \cite{knv02}.
The property is that the length is
at most $\frac{2}{2^{p(n)/2}}$ if $y\in Y_n$ and at most 2 if $y\in X_n$.
Using this property more carefully, we have a more tight bound of $l(S,T)$
as follows:
\begin{eqnarray*}
l(S,T) & = & |J_{n}\ket{\psi(S,T)}-\ket{\psi(S,T)}|\\
& = & \frac{1}{\sqrt{|S|}}\left|
	(J_{n}-I)\left(
	\sum_{y\in Y_n\cap (S\setminus T)}\ket{y}\ket{0} 
	- \sum_{y\in Y_n\cap T}\ket{y}\ket{0} 
	+ \sum_{y\in X_n\cap (S\setminus T)}\ket{y}\ket{0} 
	- \sum_{y\in X_n\cap T}\ket{y}\ket{0}
\right)\right|\\
& \le & \frac{1}{\sqrt{|S|}}\left|
	(J_{n}-I)\left(
	\sum_{y\in Y_n\cap (S\setminus T)}\ket{y}\ket{0} 
	- \sum_{y\in Y_n\cap T}\ket{y}\ket{0}\right)\right| \\
& &	+ \frac{1}{\sqrt{|S|}}\left|
	(J_{n}-I)\left(
	\sum_{y\in X_n\cap (S\setminus T)}\ket{y}\ket{0} 
	- \sum_{y\in X_n\cap T}\ket{y}\ket{0}
\right)\right|\\
& \le & \frac{1}{\sqrt{|S|}}\left(
	\sum_{y\in Y_n\cap (S\setminus T)}
	|J_{n}\ket{y}\ket{0}-\ket{y}\ket{0}| +
	\sum_{y\in Y_n\cap T}
	|J_{n}\ket{y}\ket{0}-\ket{y}\ket{0}|\right)\\
& & 
	+ \frac{1}{\sqrt{|S|}}\left(\left|
	J_{n}\left(	\sum_{y\in X_n\cap (S\setminus T)}\ket{y}\ket{0} 
	- \sum_{y\in X_n\cap T}\ket{y}\ket{0}\right)\right| +
	\left|
	\sum_{y\in X_n\cap (S\setminus T)}\ket{y}\ket{0} 
	- \sum_{y\in X_n\cap T}\ket{y}\ket{0}\right|\right)\\
& \le & \frac{2}{2^{p(n)/2}}\frac{|S\cap Y_n|}{\sqrt{|S|}}
	+ 
	\frac{2}{\sqrt{|S|}}
	\sqrt{ (|X_n\cap (S\setminus T)| + |X_n\cap T|) }\\
& = & \frac{2}{2^{p(n)/2}}
	\frac{|S\cap Y_n|}{\sqrt{|S|}}
	+ 2\sqrt{\frac{|S\cap X_n|}{|S|}}.
\end{eqnarray*}
Let $\gamma(n)$ be the former term in the above inequality.
Then
\[ \gamma(n) = \frac{2}{2^{p(n)/2}}\frac{|S\cap Y_n|}{\sqrt{|S|}} <
\frac{2^{n+1}}{2^{p(n)/2}} <
\frac{1}{2^n} \]
and is negligible.
\end{proof}

\begin{lemma}\label{lem:keep}
Let $J_{n}\ket{\psi(S,T)}=\alpha \ket{\psi(S,T)} + \ket{\psi(S,T)^{\bot}}$,
where $\ket{\psi(S,T)}\bot \ket{\psi(S,T)^{\bot}}$. Then,
$|\ket{\psi(S,T)^{\bot}}|\le l(S,T)$.
\end{lemma}

By using Lemma \ref{lem:bound} and Lemma \ref{lem:keep},
we consider the effect of the additional applications of pseudo identity
operators to {\sf INV} in order to analyze Algorithm {\sf av-INV}.

For each $j$, we let $S_{x,j}=\{y:f(y)_{(1,2j)}=x_{(1,2j)}\}$
and $T_{x,j}=\{y:f(y)_{(1,2j+2)}=x_{(1,2j+2)}\}$. 
We assume that the state before {\sf step A.j.2} is
\[ \ket{x}_1\ket{\psi(S_{x,j},T_{x,j})}_{23} 
= \ket{x}_1\frac{2^j}{\sqrt{2^n}}
\left(\sum_{y\in S_{x,j}\setminus T_{x,j}}\ket{y}_2 -
	\sum_{y\in T_{x,j}}\ket{y}_2\right)\ket{0}_3. \]
Note that the above state is the same as the one before W.$j$.2 in Algorithm {\sf INV}.

In {\sf step A.j.2}, $J_n$ is applied to the state.
From Lemma \ref{lem:bound} and a probabilistic argument,
we have the following. 

\begin{lemma}\label{lem:eval}
For each $j$, 
\[ {\bf E}[l(S_{x,j},T_{x,j})] \le \frac{2}{\sqrt{p(n)}}+\gamma(n), \]
where the expectation is over $x\in\{0,1\}^n$ and $\gamma(n)$ is a 
negligible function in $n$.
\end{lemma}
\begin{proof}
Since $f$ is a permutation, by the definition of $S_{x,j}$,
$|S_{x,j}| = 2^{n-2j}$. 
Also, $y\in S_{x,j}$ for some $x$ if and only if $y_{(1,2j)}=x_{(1,2j)}$. Then, 
\[ 
 \Pr\left[ y\in S_{x,j} \right] = \frac{2^{n-2j}}{2^n} 
 = \frac{1}{2^{2j}},
\]
where the probability is taken over $x\in\{0,1\}^n$ uniformly.
Since, for any $(1/2^{p(n)}, 1/p(n))$-pseudo identity,
\[ 
 {\bf E}[|X_n \cap S_{x,j}|] = \frac{|X_n|}{2^{2j}},
 \quad |S_{x,j}| = 2^{n-2j},
 \quad\mbox{and}\quad \frac{|X_n|}{2^n} = \frac{1}{p(n)},
\]
it holds that
\[
 {\bf E}\left[\frac{|X_n \cap S_{x,j}|}{|S_{x,j}|}\right]
 = \frac{1}{p(n)},
\]
where the expectation is over $x\in\{0,1\}^n$.
By Lemma \ref{lem:bound},
\[
 {\bf E}\left[l(S_{x,j},T_{x,j})\right]
 \le 2{\bf E}\left[\sqrt{\frac{|X_n \cap S_{x,j}|}{|S_{x,j}|}}\right]
     + \gamma(n)
 \le 2\sqrt{{\bf E}\left[\frac{|X_n \cap S_{x,j}|}{|S_{x,j}|}\right]}
     + \gamma(n)
 = \frac{2}{\sqrt{p(n)}} + \gamma(n)
\]
for some negligible function $\gamma$.
\end{proof}

From Lemma \ref{lem:keep} and Lemma \ref{lem:eval}, 
we obtain a vector $v=v_1+v_2$
where $v_1/|v_1|$ is the unit vector corresponding to the state before
{\sf step W.j.2} in Algorithm {\sf INV} and 
$v_2$ is a vector of expected length at most $2/\sqrt{p(n)}$
orthogonal to $v_1$. 
(For simplicity, we neglect a negligible term $\gamma(n)$.)
The vector $v_2$ corresponds to an error that happens
when $J_{n}$ is applied before {\sf step A.j.3}.

Next, we consider the state after step {\sf A.j.3}.
We assume that the state after {\sf step A.j.3} is
\[ \ket{x}_1\ket{\psi(S_{j+1},\varnothing)}_{23} = \ket{x}_1\frac{2^j}{\sqrt{2^n}}
\left(\sum_{y\in S_{x,j+1}}\ket{y}_2\right)\ket{0}_3. \]
Note that the above state is the same as the one after {\sf step W.j.2} in Algorithm {\sf INV}.
In order to analyze the effect of the application of 
$J_{n}^{\dag}$ after {\sf step A.j.3}, we need another
lemma similar to Lemma \ref{lem:eval}.
(The proof is omitted since its proof is also similar.)

\begin{lemma}\label{lem:eval2}
For each $j$, 
\[ {\bf E}[l(S_{x,j+1},\varnothing)] \le \frac{2}{\sqrt{p(n)}} 
+ \gamma(n), \]
where the expectation is over $x\in\{0,1\}^n$ and $\gamma(n)$ is a negligible function in $n$.
\end{lemma}

By a similar argument to the above, we obtain a vector $v=v_1+v_2$
where $v_1/|v_1|$ is the unit vector corresponding to the state after
{\sf step W.j.2} in Algorithm {\sf INV} and 
$v_2$ is a vector of expected length at most $2/\sqrt{p(n)}$
orthogonal to $v_1$. 
(For simplicity, we neglect a negligible term $\gamma(n)$.)
The vector $v_2$ corresponds to an error that happens
when $J_{n}^{\dag}$ is applied after {\sf step A.j.3}.

From the above analysis, we can see that after the completion of Algorithm
{\sf av-INV} on input $x$ the final state become $v(x)=v_1(x)+v_2(x)$ 
where $v_1(x)$ is parallel to
\[ \ket{x}_1\ket{f^{-1}(x)}_2\ket{0}_3 \]
and $v_2(x)$ is a vector orthogonal to $v_1$. By Lemma \ref{lem:eval2} 
and the linearity of expectation,
we have
\[
 {\bf E}[| v_2(x) |] \le 2 \cdot \frac{n}{2} \cdot \frac{2}{\sqrt{p(n)}} 
   = \frac{2n}{\sqrt{p(n)}} \le \frac{1}{q^2(n)}
\]
for $q(n) = p^{1/4}(n)/\sqrt{2n}$, where the expectation is over $x\in\{0,1\}^n$.
It follows that the number of $x$ such that $|v_2(x)| > 1/q(n)$ 
is at most $2^n/q(n)$, i.e., {\sf av-INV} can invert $f(x)$ 
for at least $2^n(1-1/q(n))$ $x$'s with probability at least $1-1/q^2(n)$.
\end{proofof}

\end{proof}

\section{Conclusion}
By giving a proof of the conjecture left by Kashefi, Nishimura and Vedral
\cite{knv02}, we have completed a necessary and sufficient condition of
cryptographic quantum one-way permutations in terms of pseudo-identity 
and reflection operator in this paper.

The necessary and sufficient condition of
quantum one-way permutations can be regard as
a universal test for the quantum one-wayness of permutations.
As long as the authors know, this is, classical or quantum, 
the first result on the universality for one-way permutations, 
though the next bit test is a universal test for pseudorandom generators
in the classical computation. 
We believe that our universal test
for quantum one-way permutations
may help to find good candidates for them, which are currently not known.

\subsubsection*{Acknowledgments.}
We are grateful for valuable comments from anonymous referees.
AK would like to acknowledge the financial support of
the 21st COE for Research and Education of Fundamental Technologies in
Electrical and Electronic Engineering,
Kyoto University.


\begin{thebibliography}{99}
\bibitem{ac02}
M. Adcock and R. Cleve,
``A quantum Goldreich-Levin theorem with cryptographic applications'',
In {\em Proc. 19th Annual Symposium on Theoretical Aspects of Computer Science},
Lecture Notes in Computer Science 2285,
Springer, pp.323--334, 2002.

\bibitem{bbbv97}
C. H. Bennett, E. Bernstein, G. Brassard and U. V. Vazirani,
``Strengths and weaknesses of quantum computing'',
SIAM Journal on Computing 26(5), pp.1510--1523, 1997.

\bibitem{bm84}
M. Blum and S. Micali,
``How to generate cryptographically strong sequences of
						  pseudo-random bits'',
SIAM Journal on Computing 13(4), pp.850--864, 1984.

\bibitem{bhmt00}
G. Brassard, P. H\o yer, M. Mosca and A. Tapp,
``Quantum amplitude amplification and estimation'', 
In, S. J. Lomonaco,\ Jr.\ and H. E. Brandt (eds.), {\em Quantum Computation
and Quantum Information}, AMS Contemporary Mathematics 305, AMS, 2002.

\bibitem{dms00}
P. Dumais, D. Mayers and L. Salvail,
``Perfectly concealing quantum bit commitment from
any one-way permutations'',
In {\em Advances in Cryptology --- EUROCRYPT 2000},
Lecture Notes in Computer Science 1807, 
Springer, pp.300--315, 2000.

\bibitem{gl89}
O. Goldreich and L. A. Levin,
``A hard-core predicate for all one-way functions'',
In {\em Proc. 21st ACM Symposium on Theory of Computing}, pp.25--32, 1989.

\bibitem{goldreich}
O. Goldreich,
{\em Foundations of Cryptography: Basic Tools}, Cambridge University Press,
2001.

\bibitem{gro96}
L. K. Grover,
``A fast quantum mechanical algorithm for database search'',
In {\em Proc. 28th ACM Symposium on Theory of Computing}, pp.212--219, 1996.

\bibitem{hill99}
J. H{\aa}stad,  R. Impagliazzo, L. A. Levin and M. Luby,
``A pseudorandom generator from any one-way function'',
SIAM Journal on Computing 28(4), pp.1364--1396, 1999.

\bibitem{knv02}
E. Kashefi, H. Nishimura and V. Vedral,
``On quantum one-way permutations'',
Quantum Information and Computation 2(5), pp.379--398, 2002.

\bibitem{ss93}
A. W. Schrift and A. Shamir,
``Universal tests for nonuniform distributions'',
Journal of Cryptology 6(3), pp.119--133, 1993.

\bibitem{shor97}
P. W. Shor,
``Polynomial-time algorithms for prime factorization and discrete logarithms
on a quantum computer'',
SIAM Journal on Computing 26(5), pp.1484--1509, 1997.

\bibitem{yao82}
A. C. Yao,
``Theory and applications of trapdoor functions'',
In {\em Proc. 23rd IEEE Symposium on Foundations of Computer Science}, 
pp.80--91, 1982.

\end{thebibliography}
\end{document}